\documentclass[11pt]{article}
\usepackage[utf8]{inputenc}
\usepackage[T1]{fontenc}
\usepackage[margin=1in]{geometry}
\usepackage{graphicx}
\graphicspath{{../results/}{./results/}{./}}
\usepackage{booktabs}
\usepackage{amsmath,amssymb}
\usepackage{authblk}
\usepackage[hidelinks]{hyperref}

\newcommand{\herm}{^{\mathsf{H}}}
\newcommand{\T}{^{\mathsf{T}}}
\newcommand{\Rmat}{\mathbf{R}}
\newcommand{\xv}{\mathbf{x}}
\newcommand{\av}{\mathbf{a}}
\newcommand{\wv}{\mathbf{w}}
\newcommand{\pv}{\mathbf{p}}
\newcommand{\uv}{\mathbf{u}}
\newcommand{\Ex}{\mathbb{E}}
\DeclareMathOperator*{\argmax}{arg\,max}
\DeclareMathOperator*{\argmin}{arg\,min}

\title{EchoHawk: A Reproducible Acoustic Pipeline for Drone Detection,
Classification, and Direction-Finding, with a Cautionary Study of
Session-Level Data Leakage}
\author{David Shulman\thanks{Email: david.shulman.research@gmail.com}}
\affil{\small\texttt{github.com/shulm/echohawk}}
\date{\today}

\begin{document}
\maketitle

\begin{abstract}
Passive acoustic sensing is an attractive modality for counter--unmanned-aerial-system
(counter-UAS) defence: it is covert, inexpensive, and effective against drones
with small radar cross-sections or negligible radio emissions. We present
EchoHawk, an open and fully reproducible reference pipeline that (i) detects a
drone from the harmonic structure of its rotor noise, (ii) estimates its
blade-passing frequency, and (iii) localises it with a microphone array using
classical wideband beamforming (delay-and-sum, MVDR, MUSIC) and time-delay
processing (GCC-PHAT, SRP-PHAT), followed by recursive Bayesian tracking. We
develop the underlying signal models and estimators in full, and evaluate on a
physically transparent synthetic benchmark---which deliberately pits drones
against \emph{hard} low-frequency harmonic confusers (ground vehicles)---and on
real recorded audio. Our central methodological contribution is a documented and
quantified case of \emph{session-level data leakage} in a widely used public
dataset: because its recordings are pre-segmented into short clips, naive
clip-level cross-validation places adjacent slices of the \emph{same} continuous
recording in both the training and test partitions, yielding optimistic
estimates of generalisation. Enforcing recording-session--grouped
cross-validation reduces, for example, a random-forest baseline's probability of
detection at a $5\%$ false-alarm rate from $0.937$ to $0.875$, and we report the
corrected numbers---from repeated recording-grouped cross-validation with
confidence intervals---throughout. All code, a synthetic data generator, unit
tests, and figures are released; the synthetic results reproduce with no download,
while the real-audio results require the public DroneAudioDataset.
\end{abstract}

\section{Introduction}
The accessibility and growing capability of small multirotor unmanned aerial
vehicles (UAVs) have created an urgent need for affordable, reliable
counter-UAS sensing. Each sensing modality has characteristic strengths and
failure modes. Radar offers long range and direct ranging but is challenged by
the small radar cross-section of consumer drones and by ground clutter at low
grazing angles. Electro-optical and infrared cameras provide intuitive imagery
but require an unobstructed line of sight and adequate illumination.
Radio-frequency (RF) interception of the command-and-control link is effective
against remotely piloted drones but fails against autonomous platforms that
radiate little. \emph{Acoustic} sensing is complementary to all three: a
multirotor's propellers radiate a strong and highly structured acoustic
signature whenever the vehicle is airborne, the sensor is entirely passive and
therefore difficult to detect or jam, and microphone hardware is inexpensive.
Its principal limitations---reduced range relative to radar and sensitivity to
wind and broadband environmental noise---argue for its use as one layer of a
multi-sensor system rather than in isolation.

This paper accompanies EchoHawk, an open reference implementation written with a
deliberate emphasis on \emph{methodological correctness and honest measurement}
rather than on a single headline figure. We make the following contributions.

\begin{enumerate}
\item \textbf{An open, end-to-end acoustic pipeline.} We implement and document
detection, blade-passing-frequency estimation, microphone-array direction-of-arrival
(DOA) estimation, and temporal tracking, and we release a physically transparent
synthetic generator so that every reported result reproduces with no external
download.
\item \textbf{A hard synthetic benchmark.} Rather than separating drones from
trivially dissimilar sounds, our simulator pits them against ground-vehicle
confusers whose low-frequency harmonic content overlaps that of drones, across a
wide signal-to-noise ratio (SNR) range, yielding a realistic detection problem.
\item \textbf{A quantified data-leakage study.} We identify, formalise, and
correct a session-level leakage pathology in a common public dataset, and report
the magnitude of the resulting inflation.
\item \textbf{A comparative study} of a hand-crafted-feature baseline against a
convolutional neural network (CNN) for detection, and of classical DOA estimators
as a function of SNR and array geometry.
\end{enumerate}

The remainder of the paper is organised as follows. Section~\ref{sec:related}
reviews related work. Section~\ref{sec:model} develops the acoustic and array
signal models. Section~\ref{sec:features} derives the feature representations.
Sections~\ref{sec:detection} and~\ref{sec:doa} present detection and
direction-finding. Section~\ref{sec:track} describes the tracker.
Sections~\ref{sec:data}--\ref{sec:results} cover datasets, experimental protocol,
and results, including the leakage analysis. Sections~\ref{sec:discuss}
and~\ref{sec:conc} discuss limitations and conclude.

\section{Related work}
\label{sec:related}
\paragraph{Acoustic drone detection.} Multirotor acoustic signatures are
dominated by tonal components at the blade-passing frequency (BPF) and its
harmonics, superimposed on broadband rotor turbulence. Detection systems
typically transform short audio frames into time--frequency or cepstral
representations---spectrograms, log-mel spectrograms, or mel-frequency cepstral
coefficients (MFCCs)---and apply either classical classifiers (support vector
machines, random forests, hidden Markov models) or deep neural networks
(convolutional and recurrent architectures). A recurring practical difficulty is
the discrimination of drones from other low-frequency harmonic sources such as
ground vehicles and generators, which our synthetic benchmark targets explicitly.

\paragraph{Array processing and DOA.} Estimating the direction of a sound source
from a microphone array is a classical problem. Beamforming methods range from
the non-adaptive delay-and-sum (Bartlett) beamformer to the adaptive minimum-variance
distortionless-response (MVDR/Capon) beamformer~\cite{capon}, while
subspace methods such as MUSIC~\cite{schmidt} and ESPRIT achieve
super-resolution by exploiting the eigenstructure of the spatial covariance
matrix. Time-difference-of-arrival (TDOA) approaches estimate inter-microphone
delays, most robustly through the generalised cross-correlation with phase
transform (GCC-PHAT)~\cite{knapp}, and aggregate them with the steered-response
power (SRP-PHAT) method for robust localisation in reverberant, noisy conditions.

\paragraph{Datasets.} Public resources for drone sensing include the DREGON
dataset of UAV-embedded microphone-array recordings with ground-truth source
directions~\cite{dregon}, the DroneRF dataset of radio-frequency
signatures~\cite{dronerf}, and recorded acoustic corpora such as the
DroneAudioDataset~\cite{alemadi}. We use the last for our real-audio evaluation
and provide a loader for DREGON.

\paragraph{Data leakage.} The phenomenon of information from the evaluation set
contaminating training---data leakage---is a leading cause of optimistic and
non-reproducible machine-learning results~\cite{kapoor}. We present a concrete
acoustic instance arising from the pre-segmentation of continuous recordings, and
the grouped cross-validation that remedies it.

\section{Signal and array models}
\label{sec:model}

\subsection{Acoustic model of a multirotor}
The dominant periodic excitation of a rotor with $N_b$ blades turning at
$\Omega$ revolutions per minute is the blade-passing frequency
\begin{equation}
f_0 \;=\; N_b\,\frac{\Omega}{60}\quad[\mathrm{Hz}].
\label{eq:bpf}
\end{equation}
Because the excitation is periodic, the radiated pressure contains a
\emph{harmonic stack} at integer multiples of $f_0$. We model a single-channel
drone clip $s(t)$ as a sum of harmonics with geometrically decaying amplitudes,
a slow frequency modulation that captures revolutions-per-minute jitter, and an
additive broadband term $\eta(t)$ representing rotor turbulence,
\begin{equation}
s(t) \;=\; \sum_{k=1}^{K} \rho^{\,k-1}\,
\sin\!\Big(2\pi k \!\int_0^t\! f_0(\tau)\,d\tau + \varphi_k\Big)
\;+\; \eta(t),
\qquad
f_0(t) = f_0\big(1 + \beta \sin 2\pi f_m t\big),
\label{eq:drone}
\end{equation}
where $\rho\in(0,1)$ is the harmonic decay, $\varphi_k$ random phases, and
$\beta,f_m$ the modulation depth and rate. The observation is corrupted by
additive noise $n(t)$ at a prescribed SNR,
$\mathrm{SNR}_{\mathrm{dB}} = 10\log_{10}(\sigma_s^2/\sigma_n^2)$.
The hard negative class---a ground vehicle or generator---is modelled by the same
harmonic form~\eqref{eq:drone} but with a lower fundamental, faster amplitude
decay (fewer effective harmonics), and slower modulation, so that the drone and
confuser distributions \emph{overlap} rather than separate trivially.

\subsection{Propagation}
For a point source in a homogeneous medium the pressure amplitude decays as the
inverse of range $r$ (spherical spreading), giving a sound-pressure-level
reduction
\begin{equation}
\Delta L(r) = 20\log_{10}\!\frac{r}{r_0} + \alpha(f)\,(r-r_0),
\end{equation}
i.e.\ $6$\,dB per doubling of distance plus a frequency-dependent atmospheric
absorption term $\alpha(f)$ that grows with frequency. Consequently the higher
harmonics attenuate first, and long-range detection relies on the lower members
of the stack---an observation that also informs the choice of analysis band.

\subsection{Microphone-array model}
Consider $M$ microphones at planar positions $\pv_m\in\mathbb{R}^2$ and a
far-field source whose wavefront arrives from azimuth $\theta$ with unit
direction $\uv(\theta)=(\cos\theta,\sin\theta)\T$. Relative to the array origin,
the propagation delay at microphone $m$ is
\begin{equation}
\tau_m(\theta) = -\,\frac{\pv_m\T\uv(\theta)}{c},
\end{equation}
with $c\approx343\,\mathrm{m/s}$ the speed of sound. The time-domain signal is
$x_m(t)=s\big(t-\tau_m(\theta)\big)+n_m(t)$, and in the frequency domain, using
the time-shift property,
\begin{equation}
X_m(f) = S(f)\,e^{-j2\pi f \tau_m(\theta)} + N_m(f).
\end{equation}
Collecting the channels into $\xv(f)=[X_1(f),\dots,X_M(f)]\T$ gives
$\xv(f) = S(f)\,\av(\theta,f) + \mathbf{n}(f)$, where the \emph{steering vector}
\begin{equation}
\big[\av(\theta,f)\big]_m = \exp\!\Big(j2\pi f\,\frac{\pv_m\T\uv(\theta)}{c}\Big)
\label{eq:steer}
\end{equation}
encodes the array's response to a unit-amplitude plane wave from $\theta$.
Two design rules follow from~\eqref{eq:steer}. With wavelength $\lambda=c/f$, the
inter-element spacing must satisfy $d\le\lambda/2$ to avoid spatial aliasing
(grating lobes), and the angular resolution improves with array aperture $L$
roughly as $\Delta\theta\approx\lambda/L$. An $M$-element array provides up to
$10\log_{10}M$\,dB of array gain against spatially white noise.

\section{Feature extraction}
\label{sec:features}
\subsection{Short-time Fourier transform}
All representations derive from the short-time Fourier transform (STFT). For a
window $w$ of length $N$ and hop $H$,
\begin{equation}
X(m,k) = \sum_{n=0}^{N-1} x(n+mH)\,w(n)\,e^{-j2\pi kn/N},
\qquad
S(m,k) = |X(m,k)|^2,
\end{equation}
with $S$ the spectrogram (power as a function of frame $m$ and bin $k$).

\subsection{Log-mel spectrogram and MFCCs}
A bank of $B$ triangular filters spaced uniformly on the mel scale
$\mathrm{mel}(f)=2595\log_{10}(1+f/700)$ maps the linear spectrum to perceptual
sub-bands. With $H_b(k)$ the $b$-th filter, the mel energies and their logarithm
are
\begin{equation}
E_b(m) = \sum_{k} H_b(k)\,S(m,k), \qquad
\tilde{E}_b(m) = \log\!\big(E_b(m)+\epsilon\big),
\end{equation}
and the $\tilde{E}_b$ form the \emph{log-mel spectrogram} used as the CNN input.
A discrete cosine transform decorrelates them into mel-frequency cepstral
coefficients,
\begin{equation}
c_i(m) = \sum_{b=1}^{B}\tilde{E}_b(m)\,
\cos\!\Big[\frac{\pi i}{B}\big(b-\tfrac12\big)\Big],
\qquad i=0,\dots,Q-1,
\end{equation}
which feed the classical baseline.

\subsection{Blade-passing-frequency estimation}
We estimate $f_0$ with the harmonic product spectrum (HPS), which multiplies
frequency-compressed copies of the magnitude spectrum so that energy aligned with
a true fundamental and its harmonics reinforces:
\begin{equation}
\mathrm{HPS}(f) = \prod_{h=1}^{H} \big|X(hf)\big|,
\qquad
\hat{f}_0 = \argmax_{f\in[f_{\min},f_{\max}]} \mathrm{HPS}(f).
\label{eq:hps}
\end{equation}
On a synthetic clip with $f_0=110$\,Hz,~\eqref{eq:hps} returns $109$\,Hz.

\subsection{Summary feature vector}
The baseline operates on a fixed-length vector concatenating the per-coefficient
mean and standard deviation of the MFCCs with three spectral shape descriptors
and the HPS estimate. With normalised power spectrum $P(f)=S(f)/\sum_f S(f)$, the
spectral centroid, spread, and flatness are
\begin{equation}
\mu = \sum_f f\,P(f),\quad
\sigma = \sqrt{\sum_f (f-\mu)^2 P(f)},\quad
\mathrm{SFM} = \frac{\exp\!\big(\frac{1}{F}\sum_f \log P(f)\big)}
{\frac{1}{F}\sum_f P(f)},
\end{equation}
where a low spectral flatness $\mathrm{SFM}$ indicates a tonal (harmonic) signal.

\section{Detection and classification}
\label{sec:detection}
\subsection{Detection as a hypothesis test}
Detection is the binary test between $\mathcal{H}_0$ (no drone) and
$\mathcal{H}_1$ (drone). A decision rule is characterised by its probability of
detection and probability of false alarm,
\begin{equation}
P_d = \Pr(\text{decide }\mathcal{H}_1\mid\mathcal{H}_1),\qquad
P_{fa} = \Pr(\text{decide }\mathcal{H}_1\mid\mathcal{H}_0).
\end{equation}
Sweeping the decision threshold traces the receiver operating characteristic
(ROC), and the area under it,
$\mathrm{AUC}=\int_0^1 P_d\,dP_{fa}$, summarises performance independently of any
single operating point. Because operational counter-UAS systems tolerate few
false alarms, we additionally report $P_d$ at fixed $P_{fa}\in\{1,5,10\}\%$. The
Neyman--Pearson lemma motivates thresholding a likelihood ratio; in practice we
threshold a learned class-posterior estimate, and a constant-false-alarm-rate
(CFAR) threshold can adapt to a varying noise floor.

\subsection{Classifiers}
The baseline is a random forest---an ensemble of $T$ decision trees whose
bootstrap-aggregated votes estimate the posterior---applied to the summary
feature vector. The deep model, DroneCNN, is a compact convolutional network
applied to $1$-second log-mel spectrograms; each layer computes
$\mathbf{z}^{(l)} = \phi\big(\mathbf{W}^{(l)} * \mathbf{z}^{(l-1)} + \mathbf{b}^{(l)}\big)$
with $\phi$ a rectified-linear nonlinearity and $*$ a $2$-D convolution, followed
by global pooling and a linear classifier. Given the class imbalance (drones are
the minority), the network is trained by minimising a class-weighted
cross-entropy
\begin{equation}
\mathcal{L} = -\frac{1}{n}\sum_{i=1}^{n} w_{y_i}\,
\log p_{\boldsymbol{\vartheta}}(y_i\mid \mathbf{X}_i),
\qquad w_c \propto \frac{1}{n_c},
\end{equation}
where $n_c$ is the number of training examples of class $c$, using the Adam
optimiser with weight decay and early stopping on a held-out validation split.

\section{Direction-of-arrival estimation}
\label{sec:doa}
\subsection{Spatial covariance}
All estimators below are built from the per-frequency spatial covariance matrix,
estimated by averaging outer products over $T$ STFT frames,
\begin{equation}
\Rmat(f) = \Ex\big[\xv(f)\xv(f)\herm\big]
\;\approx\; \frac{1}{T}\sum_{t=1}^{T} \xv_t(f)\,\xv_t(f)\herm
\;+\; \gamma\,\mathrm{tr}\!\big(\hat{\Rmat}(f)\big)\,\mathbf{I},
\end{equation}
where the final diagonal-loading term ($\gamma\ll1$) regularises the inverse.

\subsection{Bartlett and MVDR beamformers}
The delay-and-sum (Bartlett) beamformer steers by $\wv=\av/M$ and reports the
output power
\begin{equation}
P_{\mathrm{B}}(\theta,f) = \frac{\av(\theta,f)\herm \Rmat(f)\,\av(\theta,f)}
{\av(\theta,f)\herm\av(\theta,f)} .
\end{equation}
The MVDR/Capon beamformer instead \emph{minimises} total output power subject to
unit gain in the look direction,
\begin{equation}
\wv_{\mathrm{MVDR}} = \argmin_{\wv}\ \wv\herm\Rmat\,\wv
\quad\text{s.t.}\quad \wv\herm\av = 1,
\end{equation}
whose closed-form solution, via a Lagrange multiplier, is
$\wv_{\mathrm{MVDR}} = \Rmat^{-1}\av/(\av\herm\Rmat^{-1}\av)$, giving the spectrum
\begin{equation}
P_{\mathrm{MVDR}}(\theta,f) = \frac{1}{\av(\theta,f)\herm \Rmat(f)^{-1}\av(\theta,f)} .
\end{equation}
By suppressing energy from off-look directions, MVDR achieves a narrower main
lobe than Bartlett at the cost of sensitivity to covariance estimation error.

\subsection{MUSIC}
For $D$ sources ($D<M$), the eigendecomposition
$\Rmat = \sum_{i=1}^{M}\lambda_i\,\mathbf{e}_i\mathbf{e}_i\herm$ separates into a
signal subspace spanned by the $D$ dominant eigenvectors and a noise subspace
$\mathbf{E}_n=[\mathbf{e}_{D+1},\dots,\mathbf{e}_M]$ orthogonal to the true
steering vectors. The MUSIC pseudospectrum exploits this orthogonality,
\begin{equation}
P_{\mathrm{MUSIC}}(\theta,f) =
\frac{1}{\av(\theta,f)\herm \mathbf{E}_n\mathbf{E}_n\herm \av(\theta,f)},
\label{eq:music}
\end{equation}
exhibiting sharp peaks at the source directions.

\subsection{Wideband fusion}
Because the drone signature is broadband, we fuse narrowband spectra
incoherently over a set $\mathcal{F}$ of bins spanning the harmonic band,
\begin{equation}
\bar{P}(\theta) = \frac{1}{|\mathcal{F}|}\sum_{f\in\mathcal{F}} P(\theta,f),
\qquad
\hat{\theta} = \argmax_{\theta}\ \bar{P}(\theta).
\end{equation}

\subsection{Time-delay processing}
For a microphone pair $(i,j)$ the GCC-PHAT cross-correlation whitens the
cross-spectrum before the inverse transform, emphasising delay structure over
spectral colour,
\begin{equation}
R_{ij}(\tau) = \int \frac{X_i(f)\,X_j^{*}(f)}{\big|X_i(f)\,X_j^{*}(f)\big|}\,
e^{\,j2\pi f\tau}\,df,
\qquad
\hat{\tau}_{ij} = \argmax_{\tau} \big|R_{ij}(\tau)\big|.
\label{eq:gcc}
\end{equation}
Steered-response power (SRP-PHAT) sums the pairwise correlations evaluated at the
delays a candidate direction would induce,
$\tau_{ij}(\theta) = -(\pv_i-\pv_j)\T\uv(\theta)/c$,
\begin{equation}
P_{\mathrm{SRP}}(\theta) = \sum_{i<j} R_{ij}\!\big(\tau_{ij}(\theta)\big),
\end{equation}
yielding a robust azimuth map. A single array yields a bearing; several spatially
separated arrays/nodes intersect their TDOA hyperbolae to triangulate a position.

\section{Tracking}
\label{sec:track}
Successive single-frame bearings are noisy, but a drone's motion is smooth, so we
filter the azimuth sequence with a constant-velocity Kalman filter. With state
$\boldsymbol{\xi}=[\theta,\dot\theta]\T$, transition $\mathbf{F}$, process and
measurement covariances $\mathbf{Q},\mathbf{r}$, and observation
$\mathbf{H}=[1\ 0]$, the predict and update recursions are
\begin{align}
\boldsymbol{\xi}^- &= \mathbf{F}\boldsymbol{\xi}, &
\mathbf{P}^- &= \mathbf{F}\mathbf{P}\mathbf{F}\T + \mathbf{Q},\\
\mathbf{K} &= \mathbf{P}^-\mathbf{H}\T\big(\mathbf{H}\mathbf{P}^-\mathbf{H}\T + \mathbf{r}\big)^{-1}, &
\boldsymbol{\xi} &= \boldsymbol{\xi}^- + \mathbf{K}\big(z-\mathbf{H}\boldsymbol{\xi}^-\big),
\end{align}
with $\mathbf{P}=(\mathbf{I}-\mathbf{K}\mathbf{H})\mathbf{P}^-$. The gain
$\mathbf{K}$ trades measurement noise against model uncertainty; an
$\alpha$--$\beta$ filter is the fixed-gain special case.

\section{Datasets}
\label{sec:data}
\subsection{Synthetic benchmark}
The simulator instantiates~\eqref{eq:drone} for drones and the modified harmonic
form for ground-vehicle confusers, together with tonal distractors and pink/white
background noise, drawing SNRs uniformly in $[-15,+3]$\,dB. For DOA it synthesises
$M=8$ channels for a uniform circular array of radius $10$\,cm under the
plane-wave model~\eqref{eq:steer} plus sensor noise, with a known ground-truth
azimuth. Because the generative process is fully specified, every result is
reproducible from a seed with no download.

\subsection{Real audio: DroneAudioDataset}
The DroneAudioDataset~\cite{alemadi} comprises recorded drone audio and everyday
sounds (drawn from ESC-50 and a speech-commands corpus). After resampling to
$16$\,kHz and segmentation into $1$-second windows we obtain $19{,}275$ clips
($2{,}420$ drone, $16{,}855$ non-drone). Critically, the $1{,}332$ drone files
originate from only $257$ continuous recording \emph{sessions}, identifiable from
filename prefixes such as \texttt{B\_S2\_D1\_067-bebop}; the $16{,}855$ negative
clips likewise derive from a limited set of source recordings ($1{,}896$ groups
after grouping ESC-50 by source clip and Speech Commands by speaker). This
structure is central to Section~\ref{subsec:leak}.

\section{Experimental setup}
\label{sec:setup}
Audio is resampled to $16$\,kHz and segmented into $1$-second windows. The STFT
uses a Hann window. All randomness is seeded for exact reproducibility. For the
real data we use \emph{recording-grouped} cross-validation---$5$ folds for the
random forest and $3$ folds for the CNN, on identical outer folds---grouping
segments by recording session for drones and by source clip (ESC-50) or speaker
(Speech Commands) for the negatives, so that no source family crosses a fold; we
verify train/validation/test group-disjointness by assertion. Within each CNN
training fold a further group-disjoint split provides the early-stopping
validation set. We report each metric as the mean $\pm$ 95\% confidence interval
over folds: ROC-AUC, accuracy, and $P_d$ at $1$, $5$, and $10\%$ false alarm
(read from the linearly interpolated ROC), macro-averaged $F_1$ for multi-class
tasks, and root-mean-square azimuth error for DOA.

\section{Results}
\label{sec:results}
\subsection{Synthetic detection and direction-finding}
\begin{figure}[h]\centering
\includegraphics[width=0.70\linewidth]{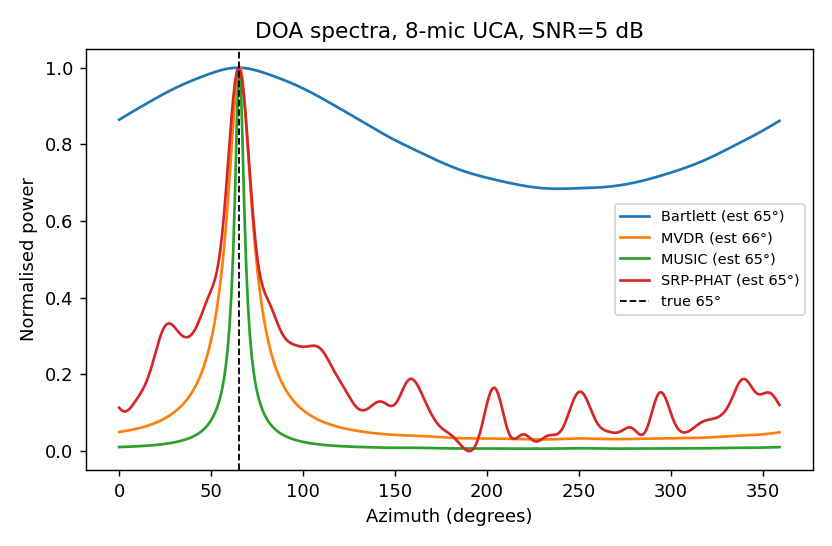}
\caption{Azimuth spectra for a source at $65^\circ$ (SNR $=5$\,dB). All four
estimators peak at the true bearing; MUSIC is sharpest, Bartlett broadest---a
direct illustration of the resolution/robustness trade-off.}
\label{fig:doa}
\end{figure}

On the synthetic hard-confuser task the random-forest baseline attains
$\mathrm{AUC}\approx0.92$ and accuracy $\approx0.86$ under $5$-fold cross-validation; the deliberately overlapping
drone and vehicle distributions make this a realistic, non-trivial problem rather
than a separable toy. Table~\ref{tab:doa} reports DOA root-mean-square error
versus array SNR for the eight-microphone array. As predicted by theory, MVDR and
MUSIC outperform delay-and-sum at low SNR, where their narrower main lobes resolve
the source from the noise floor; all methods converge to a fraction of a degree
at high SNR (Fig.~\ref{fig:doa}).

\begin{table}[h]\centering
\caption{Synthetic DOA root-mean-square error (degrees) versus array SNR,
8-microphone uniform circular array.}
\label{tab:doa}
\begin{tabular}{lccc}
\toprule
Array SNR & Bartlett & MVDR & MUSIC \\
\midrule
$-10$\,dB & $\sim$4.0 & $\sim$1.6 & $\sim$2.3 \\
$0$\,dB   & $\sim$0.6 & $\sim$0.8 & $\sim$0.9 \\
$+10$\,dB & $\sim$0.3 & $\sim$0.3 & $\sim$0.3 \\
\bottomrule
\end{tabular}
\end{table}

\subsection{Real-data detection}
Under recording-session--grouped evaluation the CNN outperforms the baseline
across all metrics, and the gap is largest in the operationally important
low-false-alarm regime (Table~\ref{tab:real}). At a $1\%$ false-alarm rate the
CNN detects $93.8\%$ of drones against $72.3\%$ for the baseline, reflecting the
benefit of learning the full spectro-temporal pattern rather than a hand-designed
summary. Because both the drone (session) and the negative (source) classes are
group-disjoint across folds, this is a leakage-free estimate for both classes.

\begin{table}[h]\centering
\caption{Real-data detection under recording+source-grouped cross-validation
($5$-fold RF, $3$-fold CNN; mean $\pm$ 95\% CI).}
\label{tab:real}
\begin{tabular}{lcc}
\toprule
Metric & RF baseline & DroneCNN \\
\midrule
ROC-AUC        & $0.966\pm0.020$ & $\mathbf{0.991\pm0.010}$ \\
Accuracy       & $0.954\pm0.010$ & $\mathbf{0.969\pm0.012}$ \\
$P_d$ @ $1\%$ FA  & $0.723\pm0.047$ & $\mathbf{0.938\pm0.026}$ \\
$P_d$ @ $5\%$ FA  & $0.875\pm0.036$ & $\mathbf{0.968\pm0.037}$ \\
$P_d$ @ $10\%$ FA & $0.924\pm0.030$ & $\mathbf{0.975\pm0.037}$ \\
\bottomrule
\end{tabular}
\end{table}

\subsection{The session-level leakage study}
\label{subsec:leak}
We now formalise and quantify the central methodological point. Let the dataset
be $\mathcal{D}=\{(\mathbf{X}_i,y_i,g_i)\}_{i=1}^{n}$, where $g_i$ indexes the
continuous recording \emph{session} from which clip $i$ was cut. Clips sharing a
session, $g_i=g_j$, also share latent nuisance factors---the same background, the
same channel, the same drone unit at near-identical operating point. A random
clip-level split assigns such clips independently and therefore, with high
probability, places near-duplicates of one recording in both training and test.
A classifier can then exploit the session-identifying nuisance, estimating
$\Pr(y\mid g)$ rather than the intended $\Pr(y\mid \mathbf{X})$, and the empirical
test risk understates the true risk on \emph{unseen} recordings. The remedy is a
group-constrained partition with
$\mathcal{G}_{\mathrm{train}}\cap\mathcal{G}_{\mathrm{test}}=\varnothing$, which
restores an (approximately) unbiased estimate of generalisation to new recordings.

The correct grouping unit is the \emph{recording}: for drones the continuous
session (recovered from the filename prefix, giving $257$ groups), and for the
negatives the source recording---the ESC-50 source clip or the speaker---rather
than the individual segment, giving $1{,}896$ groups. Grouping only the drones,
as a partial fix would, leaves the negative class leaking through source-family
siblings; we therefore group both classes and assert fold-wise disjointness.
Table~\ref{tab:leak} contrasts file-grouped evaluation with the corrected
recording+source grouping. Sealing the leak lowers the scores, most visibly for
the random forest (its $P_d$@$5\%$FA falls from $0.937$ to $0.875$), exposing the
residual inflation. The corrected numbers in Table~\ref{tab:real} are the ones we
regard as trustworthy.

\begin{table}[h]\centering
\caption{File-grouped versus corrected recording+source-grouped metrics
($5$-fold RF, $3$-fold CNN). The drop quantifies the inflation removed by correct
grouping.}
\label{tab:leak}
\begin{tabular}{lcccc}
\toprule
 & \multicolumn{2}{c}{RF baseline} & \multicolumn{2}{c}{DroneCNN} \\
\cmidrule(lr){2-3}\cmidrule(lr){4-5}
Metric & File-grouped & Corrected & File-grouped & Corrected \\
\midrule
ROC-AUC       & 0.986 & 0.966 & 0.996 & 0.991 \\
$P_d$ @ $5\%$ FA & 0.937 & 0.875 & 0.979 & 0.968 \\
\bottomrule
\end{tabular}
\end{table}

\subsection{Tracking}
On a moving-source simulation in which the true bearing sweeps across azimuth over
roughly three seconds, the Kalman/$\alpha$--$\beta$ tracker reduces the bearing
root-mean-square error from $\sim5.7^\circ$ (raw per-frame MUSIC) to
$\sim1.9^\circ$, confirming that exploiting temporal continuity substantially
stabilises the estimate.

\section{Discussion and limitations}
\label{sec:discuss}
Several limitations bound the interpretation of these results, and we state them
explicitly. First, the real-audio benchmark uses \emph{easy} negatives---everyday
environmental and speech sounds rather than operational confusers such as ground
vehicles---so its high absolute scores overstate the difficulty of fielded
operation; the hard drone-versus-vehicle discrimination is exercised only on the
synthetic benchmark. Second, the DOA evaluation is predominantly synthetic: a
loader and runnable script for the real DREGON microphone-array dataset are
provided, but physical-array results await data access, and DREGON's UAV-mounted
geometry additionally introduces strong ego-noise. Third, the array processing is
developed for a single, far-field source; near-field effects, multiple
simultaneous sources, multipath, and strong wind are out of scope. Finally, the
synthetic generator, while physically motivated, is an idealisation. None of
these caveats affects the central methodological contribution---the leakage
analysis and the discipline of grouped evaluation---which transfers to any
study that segments continuous recordings into clips.

\section{Conclusion}
\label{sec:conc}
We presented EchoHawk, a compact, open, and fully reproducible acoustic pipeline
for drone detection, blade-passing-frequency estimation, microphone-array
direction-finding, and tracking, together with the underlying signal models and
estimators. Beyond the implementation, the most transferable result is
methodological: when short clips are cut from a small number of continuous
recordings, only recording-grouped cross-validation yields honest performance
estimates; clip- or file-level splits inflate them measurably. We release all
code, a synthetic data generator, tests, continuous integration, and figures so
that the results---and the cautionary analysis---can be reproduced and built upon.
Future work includes evaluation on real microphone-array data (DREGON), hard
real-world confusers, multi-source and multipath settings, and end-to-end fusion
of detection with localisation across networked nodes.

\end{document}